\documentclass[%
 reprint,
 amsmath,amssymb,
 aps,
]{revtex4-1}

\usepackage{graphicx}% Include figure files
\usepackage{dcolumn}% Align table columns on decimal point
\usepackage{bm}% bold math
\usepackage{braket}
\usepackage{xcolor}
\usepackage{mathtools}

\begin{document}

\newcommand{\tbox}[1]{\mbox{\tiny #1}}

\preprint{APS/123-QED}

\title{Survival probability of random networks}% Force line breaks with \\

\author{Kevin Peralta-Mart\'inez}
\email{kperaltamtz@gmail.com}
\affiliation{Departamento de F\'isica, Universidad Aut\'onoma Metropolitana-Iztapalapa, Apartado Postal 55-534, 09340 Ciudad de Mexico, Mexico.}%Lines break automatically or can be forced with \\
\author{J. A. M\'endez-Berm\'udez}%
 \email{jmendezb@ifuap.buap.mx}
\affiliation{Instituto de F\'isica, Benem\'erita Universidad Aut\'onoma de Puebla, 
Apartado Postal J-48, Puebla 72570, Mexico.}%Lines break automatically or can be forced with \\

\label{abstract}
\begin{abstract}
In this work we study in detail all phases of the time evolution of a delta-like excitation in Erd\"os-Renyi (ER) 
random networks by means of the survival probability (SP):
The initial decay of the SP (both, the fast decay followed by the power-law decay), the correlation hole regime 
(the regime between the minimum value of the SP and its saturation value), and the saturation of the SP.
Specifically, we found that 
(i) the power-law decay of the SP and the time-averaged SP are proportional to $t^{-D_{2}}$ and $t^{-\widetilde{D}_{2}}$,
respectively (where $D_2$ and $\widetilde{D}_2$ are the correlation dimension of the eigenstates of the randomly 
weighted adjacency matrices of the ER random networks and the correlation dimension associated with the initial state,
respectively) and
(ii) the relative depth of the correlation hole of the SP scales with the average degree $\langle k\rangle\approx np$
(here, $n$ and $p$ are the size and the connection probability of the ER random networks).
In addition, we show that the eigenstates of the randomly weighted adjacency matrices of ER networks display 
clear multifractal properties.
\end{abstract}

\maketitle

\section{Introduction}

The study of random networks has become crucial for the understanding of complex systems 
composed by several interconnected elements. While structural and spectral properties of 
networks have been extensively analyzed, the dynamics of processes occurring on networks 
and graphs have also attracted great interest. Specific examples are studies of: rumor 
spreading~\cite{MNP04}, 
% glassy systems modelling with random regular graphs~\cite{MKS19}, 
the return or survival probability of random walks in scale-free trees~\cite{C11}, and random 
walks described by a discrete time Markovian process~\cite{RM19}. Moreover, the use of 
fractional calculus has permitted to explore fractional diffusion in networks, allowing the study 
of random walk dynamics with long range interactions~\cite{RM14}. As an extension, fractional 
quantum transport in ring graphs was studied in Ref.~\cite{RM15}, leading to a generalized 
dynamics based on continuous-time quantum walks~\cite{MB11}. More recently, Mellin-transformed 
$d-$path Laplacian operators have been implemented to the study of quantum transport and return 
probabilities of ring, complete, and star graphs~\cite{E20}.

In this work, we aim to contribute to the understanding of quantum dynamics in random networks 
by analyzing the evolution of an excitation in Erd\"os-Renyi (ER) random networks in the framework 
of random matrix theory (RMT). 
Specifically, we focus on the survival probability of a delta-like excitation in terms of the inherent 
parameters of the ER model. This work is organized as follows.
In Sec.~\ref{Sec:Preliminaries} we introduce the ER model and the corresponding weighted 
adjacency matrix that we use as a framework for the study of the survival probability. 
Then, in Sec.~\ref{Sec:Analysis SP} we analyze in detail the different regimes in the time evolution 
of the survival probability: The decay of the survival probability (the decay after the exponential initial 
decay), the correlation hole (the regime between the minimum value of the survival probability and 
its saturation value), and the saturation of the survival probability.
In addition, in Appendix~\ref{append:Multifractal dimensions of eigenstates}, we show that the
eigenstates of the randomly weighted adjacency matrices of ER networks display multifractal
properties.
Finally, in Sec.~\ref{Sec:Conclusions} we draw our conclusions.

\section{Preliminaries} 
\label{Sec:Preliminaries}

\subsection{ Erd\"os-Renyi model} 
\label{Subsec:ErdosRenyi}

The ER random network model, introduced in 1951 by Solomonoff and Rapoport~\cite{SR51} 
and later named after Paul Erd\"os and Alfred Renyi~\cite{ER59}, is generated by independently connecting 
$n$ vertices or nodes with probability $p$ through a random pairing process. This pairing process leads to a 
graph where the presence or absence of any specific edge is entirely uncorrelated with the others. 
Thus, given the random and independent nature of edge formation, the degree $k$ of any particular node in 
an ER graph follows a binomial distribution as it is the sum of $n-1$ independent Bernoulli trials, each with 
success probability $p$. Then, the average degree can be calculated directly from the parameters $n$ and $p$. 
Since each node has $n-1$ potential connections, and each connection exists with probability $p$, the 
expected degree $\langle k\rangle$ of any node is:
\begin{equation}
\langle k\rangle = (n-1)p \approx np.
\label{eq:Averagedegree_ER}
\end{equation}
Therefore, in the case where $p=c/n$ and $c$ is a constant, the average degree shows to be independent of 
the network size $n$. Then, as $c$ increases, the network becomes more densely connected which influences 
various properties of the network, such as the emergence of a giant component and the overall robustness of 
the network. In fact the average degree is a key parameter that influences the overall structure of the network 
and connectivity and it has been shown to scale structural as well as spectral properties of various random 
network models; see e.g.~\cite{PRRCM20,MAMRP15,AMGM18,PM23,AMRS20}.

\subsection{Weighted adjacency matrix}
\label{Subsec:Weighted adjacency matrix}

The adjacency matrix has been widely used to analyze structural and spectral properties of 
diverse random network models. Specifically, with the help of binary adjacency matrices, the 
spectral properties of ER random networks have been deeply studied; see 
e.g.~\cite{EKYY13,EKYY12,PRRCM20,MA15}.
Moreover, weighted versions of the adjacency matrix have also been explored as a bridge to 
RMT ensembles and to use RMT predictions as a reference in the appropriate limits. 
As examples we can mention that Ref.~\cite{MA15} focuses on undirected ER networks, 
Ref.~\cite{PRRCM20} on directed ER networks, Refs.~\cite{AMGM18,MAMRP15} on undirected 
random geometric graphs (RGGs), and more recently, Ref.~\cite{PM23} addresses directed RGGs. 
In connection to these works, here we propose the study of the survival probability in ER random 
networks represented by the following weighted adjacency matrix:
\begin{equation}
\left[{\bf A}_{\tbox{RMT}}\right]_{ij}  \equiv \left[{\bf A}\right]_{ij} = \left\{
\begin{array}{rl}
        \sqrt{2}\epsilon_{ij}& \ \ \  \text{if $i=j$} \\ \vspace{1mm} \ \ \epsilon_{ij}&   \ \ \    \text{if $i\leftrightarrow j$}\\ \vspace{1mm}\ \ 0&\   \ \ \text{otherwise}
\end{array}
\right.,
\label{eq:Weighted_adjacency_matrix}
\end{equation}
where $\epsilon_{ij}$ are independent random variables drawn from a normal distribution with
mean 0 and variance $1$.  
Therefore, with this definition for ${\bf A}$, when the networks are mostly disconnected ($p\rightarrow0$), 
the resulting diagonal adjacency matrices are members of the Poisson ensemble (PE). On the other 
hand, adjacency matrices belonging to the Gaussian Orthogonal Ensemble (GOE) are obtained when 
the networks become fully connected as $p\rightarrow1$. Consequently, we expect to observe a transition 
in the evolution of the dynamical properties of the ER random networks by increasing $p$, moving from 
the PE regime towards the GOE limit. Hence, when we consider weighted adjacency matrices as in 
equation~(\ref{eq:Weighted_adjacency_matrix}), we deal with a diluted GOE.

\subsection{Survival probability}
\label{Subsec:Survival probability}

As already mentioned in the Introduction, in this paper we characterize the dynamics of ER random 
networks by the use of the survival probability 
(SP)~\cite{TKS16,STS19,TTS16,TTS17,TKS15,ZST23}, also known as return probability~\cite{IC06,KOY11}, 
which measures the probability of finding the system in its initial state after the time $t$. 
Assuming an initial state $\ket{\Psi(t=0)}$ and its time-evolved counterpart $\ket{\Psi(t)}$, 
the SP is the squared inner product of both.
In the context of the GOE, for the analysis of the SP, it is assumed the 
existence of a basis of states $\ket{\phi_{k}}$ used to construct every GOE matrix.
Thus, we define an initial state $\ket{\phi_{ini}}\equiv \ket{\Psi(0)}$ at the bulk of the matrix, $ini=n/2$~\cite{TKS16},
with $n$ the size of the matrix. Then the SP is defined as~\cite{TKS16}
\begin{align}
SP(t)=&|\bra{\Psi(0)}\ket{\Psi(t)}|^2 = \left|\sum_{\beta}\left|C_{ini}^{\beta}\right|^{2}e^{-iE_{\beta}t} \right|^2 \\ \notag = & \left| \int{\rho(E) e^{-iE_{\beta}t}dE} \right|^{2}, 
\label{eq:SP}
\end{align}
where 
\begin{equation}
\rho(E) = \sum_{\beta}\left|C_{ini}^{\beta}\right|^{2}\delta(E-E_{\beta}) 
\end{equation}
corresponds to the energy distribution of the initial state, also known as the local density of states (LDOS). 
Since the SP and the LDOS are related by a Fourier transform, the SP can be analytically derived once the 
LDOS shape is determined. For the GOE, the shape of the LDOS follows the Wigner semicircle law,
\begin{equation}
\rho(E)= \frac{1}{\pi\sigma_{ini}}\sqrt{1-\left(\frac{E}{2\sigma_{ini}}\right)},
\label{eq:Semicircle}
\end{equation}
where $\sigma_{ini}^{2}$~\cite{TKS16} is the variance of the energy distribution of the initial state:
\begin{equation}
{\sigma_{ini}}^2 = \sum_{\beta}\left|C_{ini}^{\beta}\right|^2 \left(E_{\beta} - \sum_{\beta}\left|C_{ini}^{\beta}\right|^2E_{\beta}\right)^2,
\label{eq:sigma}
\end{equation}
with $E_{\beta}$ being the eigenenergies of the GOE.

The analytical expression for the evolution of the SP~\cite{STS19} for the GOE
predicts fast decaying oscillations at short times as $t^{-3}$~\cite{TTS16,TTS17,TKS15} according 
to the Fourier transform of the semicircle; $[\mathcal{J}_{1}(2\sigma_{ini}t)/(\sigma_{ini} t)]^2$, 
where $\mathcal{J}_{1}$ is a Bessel function of the first kind.
We will use this result as a reference for the diluted GOE corresponding to the weighted adjacency 
matrices of ER networks.

\section{Analysis of the survival probability}
\label{Sec:Analysis SP}

\begin{figure*}[ht]
\centering
\includegraphics[width=0.75\textwidth]{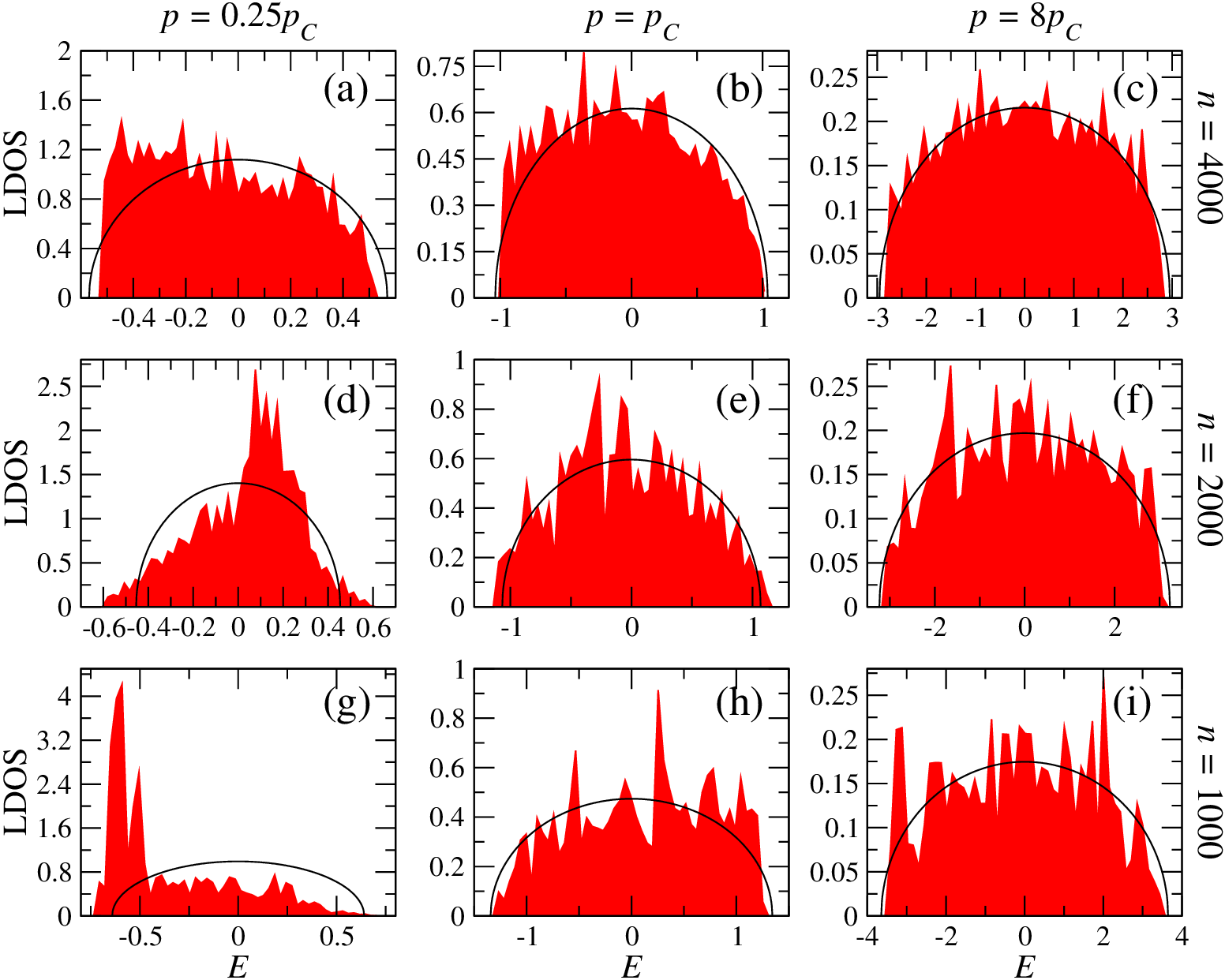}
\caption{
Histograms of the local density of states (LDOS) at the center of the band of ER random networks. 
Several combinations of sizes $n$ (different rows) and connection probabilities $p$ (different
columns) are considered. Specifically, (a,d,g) $p=0.25p_c$, (b,e,h) $p=p_c$, and (c,f,i) $p=8p_c$ with
$p_{c}=n^{2/3}/(n-1)$. Each histogram was constructed from a single random network realization. Black 
lines are the semicircles of Eq.~(\ref{eq:Semicircle}) with $\sigma_{ini}$ given by Eq.~(\ref{eq:sigma}).
}
\label{fig:fig01}
\end{figure*}

\begin{figure*}[ht]
\centering
\includegraphics[width=0.75\textwidth]{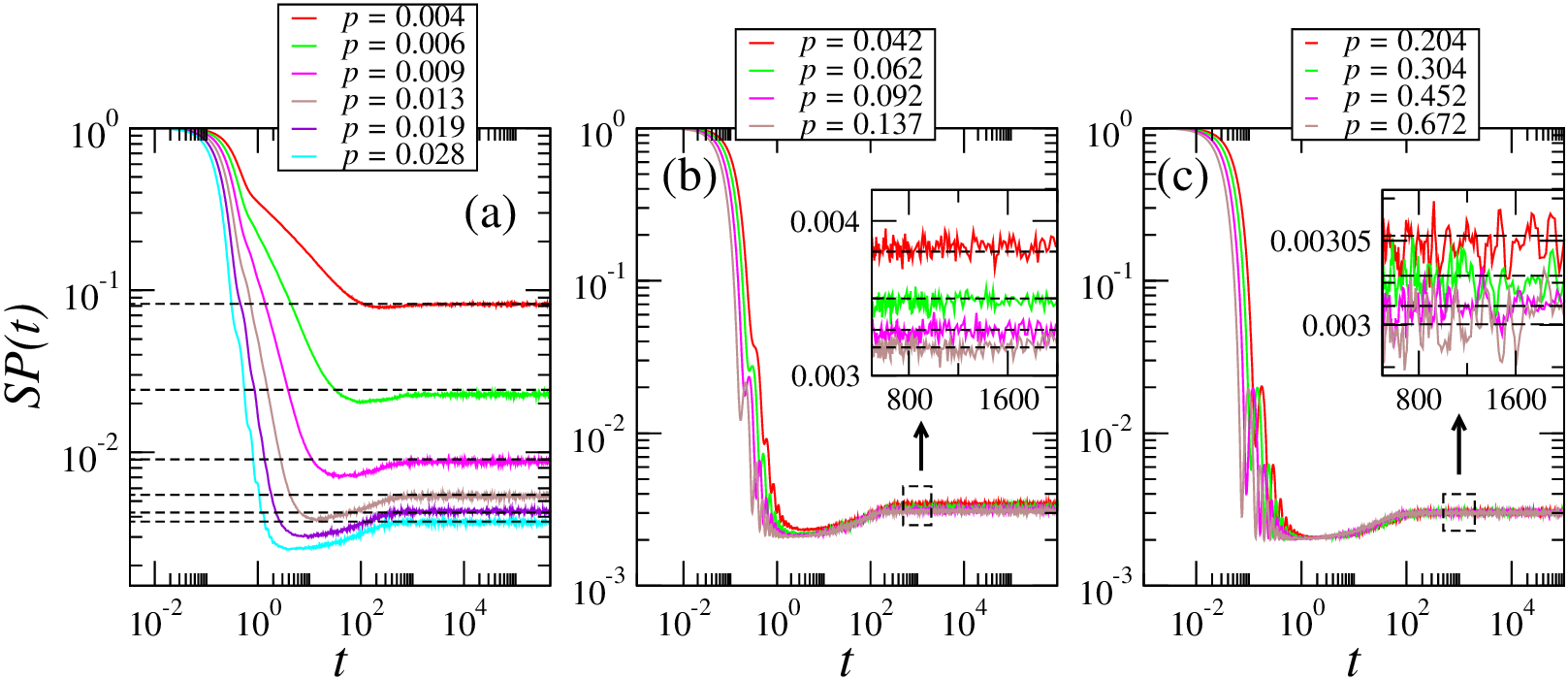}
\caption{
Survival probability $SP(t)$ of ER random networks of size $n=1000$ and several values of 
the connection probability $p$. Insets in panels (b,c) are enlargements of the corresponding dashed 
squares. Horizontal dashed black lines indicate the saturation value of the curves.
The connection probability $p$ increases from top to bottom.
}
\label{fig:fig02}
\end{figure*}

We start by exploring the LDOS of ER random networks.
In fact, in Ref.~\cite{EKYY13} it was shown that the LDOS is expected to follow the Wigner 
semicircle law (when $n\rightarrow\infty$) for $p>p_{c}$ with $p_{c}\sim ({n^{2/3}})/(n-1)$.
Thus, in Fig.~\ref{fig:fig01} we present histograms of the LDOS at the center of the band 
of ER random networks with $p<p_c$ (left panels), $p=p_c$ (central panels), and $p>p_c$ 
(right panels). Indeed, we verify that when $p>p_c$ the LDOS of ER random networks 
follows closely a semicircular shape. As a consequence, we expect to observe a GOE 
behavior of the SP~\cite{STS19} for ER random networks with $p>p_c$. In contrast, 
when $p<p_c$ the LDOS 
deviates strongly from the semicircle law. Thus, when $p<p_c$, the SP 
is expected to show significant deviations from the GOE behavior.

Now, in Fig.~\ref{fig:fig02} we present the temporal evolution of the SP for an ensemble 
of $2\times10^4$ ER random networks of size $n=1000$ and several values of the connection 
probability $p$. 
From Fig.~\ref{fig:fig02} we clearly observe the different regimes of the SP behavior~\cite{ZST23}:
The universal exponential initial decay is followed by a slower decay until the SP reaches a minimum 
value, then it shows a small recovery (ramp) before saturation.

In what follows we analyze in detail the different regimes in the time evolution of the SP:
The decay of the SP (the decay after the exponential initial decay), the correlation hole (the regime
between the minimum value of the SP and its saturation value), and the saturation of the SP.

\subsection{Decay of the SP}
\label{subsec:Decay of the SP}

\begin{figure*}[ht]
\centering
\includegraphics[width=0.75\textwidth]{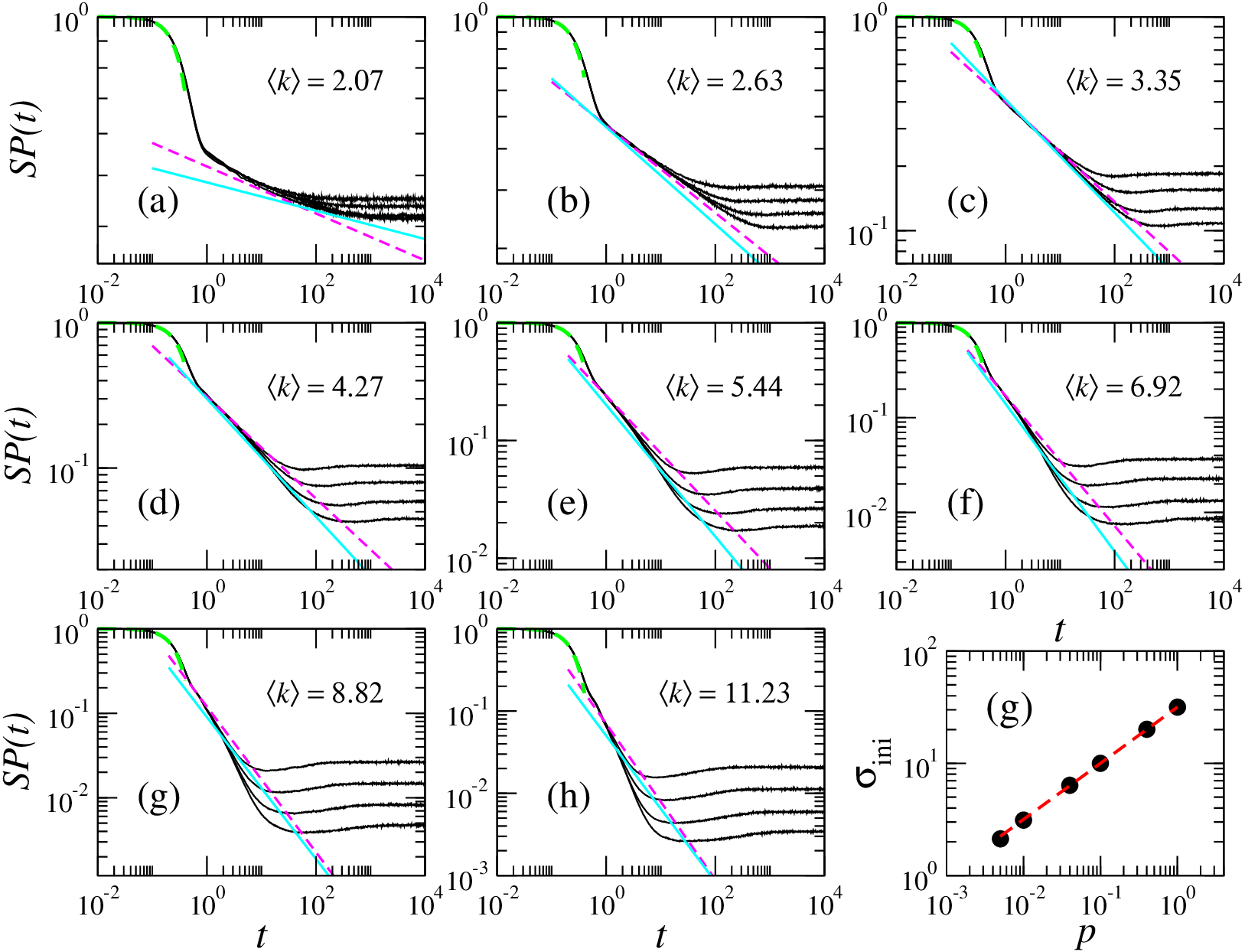}
\caption{
(a-h) Survival probability $SP(t)$ of ER random networks for several values of the average degree 
$\langle k\rangle$, as indicated in the panels. Four graph sizes are reported in each panel ($n=250$, 
500, 1000, 2000); they increase from top to bottom. 
Magenta dashed lines correspond to $t^{-D_{2}}$ with (a) $D_2=0.0446$, (b) $D_2=0.1316$, 
(c) $D_2=0.234$, (d) $D_2=0.3509$, (e) $D_2=0.4885$, (f) $D_2=0.6868$, (g) $D_2=0.8615$, and 
(h) $D_2=0.9477$. 
Cyan solid lines correspond to $t^{-\widetilde{D}_{2}}$ with (a) $\widetilde{D}_2=0.0269$, (b) $\widetilde{D}_2=0.1467$, 
(c) $\widetilde{D}_2=0.2647$, (d) $\widetilde{D}_2=0.4062$, (e) $\widetilde{D}_2=0.5589$, (f) $\widetilde{D}_2=0.7786$, 
(g) $\widetilde{D}_2=0.8398$, and (h) $\widetilde{D}_2=0.8873$. 
Green dashed lines correspond to $1-\langle k\rangle t^{2}$; i.e.~the decay at very short times. 
(g) Standard deviation of the energy distribution of the initial state $\sigma_{ini}$ (Eq.~(\ref{eq:sigma})) 
as a function of the connection probability $p$ for ER random networks of size $n=1000$. 
The red dashed line is a fit of the function $\sigma_{ini}=Ap^{B}$ to the data with $A=31.682$ and $B=0.501$.
} 
\label{fig:fig03}
\end{figure*}

\begin{figure*}[ht]
\centering
\includegraphics[width=0.75\textwidth]{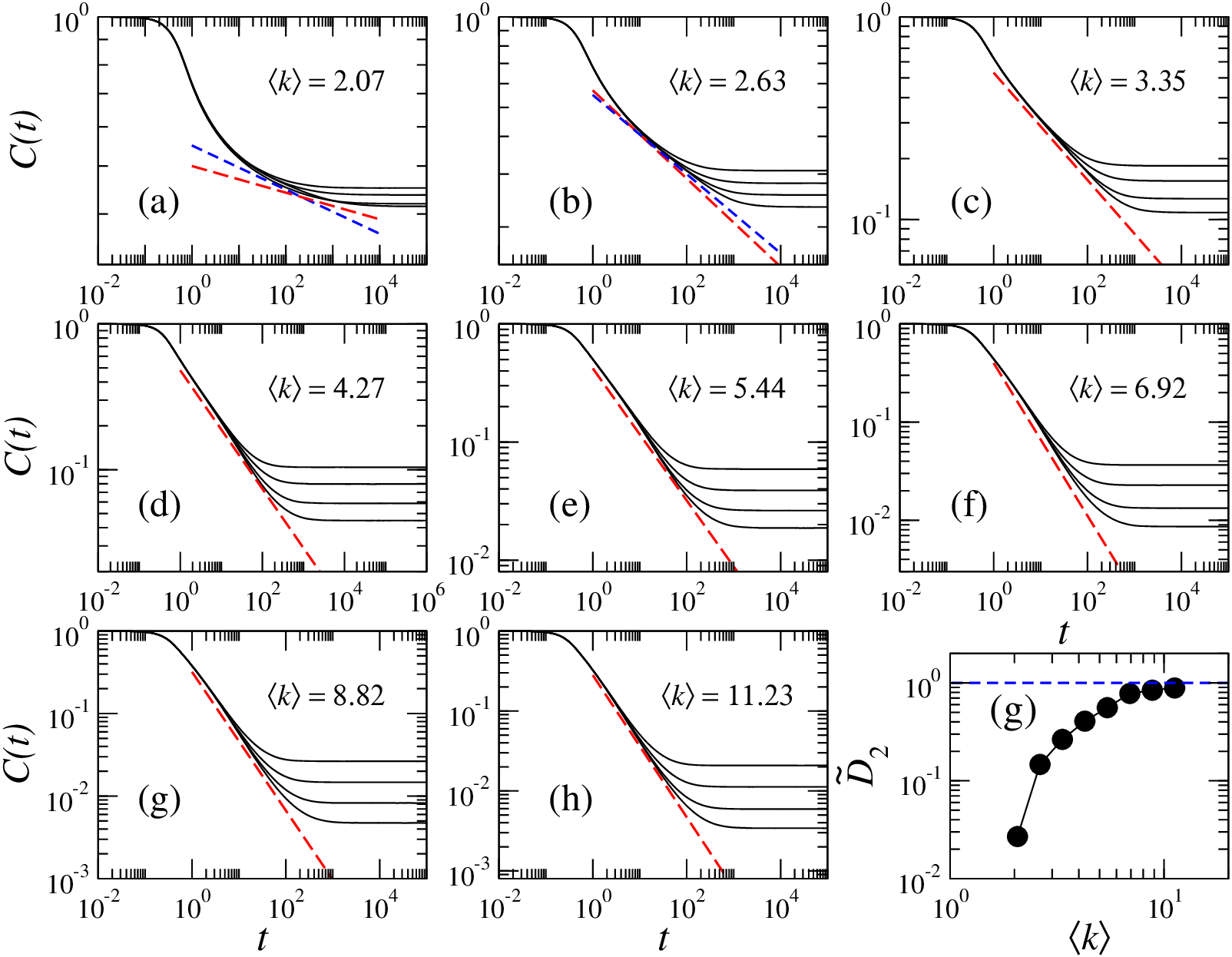}
\caption{
(a-h) Time-averaged survival probability $C(t)$ of ER random networks for several values of the average degree
$\langle k\rangle$, as indicated in the panels. Four graph sizes are reported in each panel ($n=250$, 
500, 1000, 2000); they increase from top to bottom. Red dashed lines correspond to $t^{\widetilde{D_{2}}}$.
In panel (a) the blue dashed line corresponds to $t^{D_{2}}$ and is plotted for comparison purposes. 
(g) Generalized dimension $\widetilde{D}_{2}$ of the initial state $\ket{\phi_{ini}}$ as a function of the average 
degree $\langle k\rangle$. The blue dashed line indicates $\widetilde{D_{2}}=1$.
}
\label{fig:fig04}
\end{figure*}

The initial fast decay of the SP can be well described by the dominant term in the power series expansion of the 
inner product $|\bra{\Psi(0)}\ket{\Psi(t)}|^2$~\cite{TKS16},
\begin{equation}
SP(t\ll 1) \approx 1 - \sigma^{2}_{ini}t^2 ,
\label{eq:ApproxSP}
\end{equation}
where $\sigma_{ini}$ corresponds to Eq.~(\ref{eq:sigma}). 

In order to characterize the standard deviation of the energy distribution of the initial state $\sigma_{ini}$ in 
terms of the connection probability $p$, in Fig.~\ref{fig:fig03}(g) we plot $\sigma_{ini}$ for several values of $p$ 
for networks of size $n=1000$. 
Here, we find that the $\sigma_{ini}$ vs. $p$ curve follows a power-law of the form $\sigma_{ini}= Ap^B$ with fitting 
values $A = 31.68175\approx 1000^{1/2}$ and $B = 0.5006\approx 1/2$. This shows that $\sigma_{ini}$ is directly 
related to the average degree $\langle k \rangle$ as $\sigma_{ini}\approx \sqrt{np}\approx \langle k\rangle ^{1/2}$; 
see Eq.~(\ref{eq:Averagedegree_ER}). Thus, in Figs.~\ref{fig:fig03}(a-h) we plot
\begin{equation}
SP(t\ll 1)\approx 1-\langle k\rangle t^2 
\label{SPfast}
\end{equation}
(as green dashed lines) and observe a very good correspondence with the SP curves at small times for different 
values of $\langle k \rangle$.

Now, from Fig.~\ref{fig:fig02}(a), when $p<p_c$ (here $p_{c}=0.\overline{11}$), we observe a clean power-law 
decay,
\begin{equation}
SP(t)\sim t^{-\mu},
\label{mu}
\end{equation}
of the SP after the initial fast decay. 
When $p\to p_c$, see Fig.~\ref{fig:fig02}(b), oscillations emerge along the decay of the SP.
This oscillations increase in amplitude for increasing $p$, see Fig.~\ref{fig:fig02}(c), approaching the
GOE line-shape prediction for the SP~\cite{TTS16,TTS17,TKS15}.

Indeed, studies of disordered noninteracting systems have shown that the power $\mu$ in the power-law 
decay of the SP, see Eq.~(\ref{mu}), is directly related to the correlation dimension $D_2$ of the 
corresponding eigenstates, i.e.~$\mu\approx D_2$~\citep{HK99,HS94,KPG92}; see 
Appendix~\ref{append:Multifractal dimensions of eigenstates}.
Also, in the study of the dynamics at the many-body localization transition, the correlation dimension 
associated with the initial state $\widetilde{D_2}$ has also been related to $\mu$,
i.e.~$\mu\approx \widetilde{D_2}$~\cite{TS15}; see also Appendix~\ref{append:Multifractal dimensions of eigenstates}.
Therefore in what follows we compute $D_2$ and $\widetilde{D_2}$ and compare them with the power
$\mu$.

Since we aim to characterize the power-law behavior of the $\mathrm{SP}$ curves and the average degree fixes 
the localization properties of the eigenstates of the network, see e.g.~\cite{AHMS20},
in Fig.~\ref{fig:fig03} we show $SP(t)$ plots for different fixed values of $\langle k\rangle$, from 
$\langle k\rangle=2.07$ [see panel (a)] to $\langle k\rangle=11.23$ [see panel (h)] and four network sizes $n$, 
increasing from top to bottom. Dashed-magenta and cyan lines proportional to $t^{-D_{2}}$ and $t^{-\widetilde{D}_{2}}$, 
respectively, are shown in Fig.~\ref{fig:fig03} to guide the eye. 
From Fig.~\ref{fig:fig03}(a) we note that neither $t^{-D_{2}}$ nor $t^{-\widetilde{D}_{2}}$ match the decay of $SP(t)$.
However as $\langle k\rangle$ increases, $t^{-D_{2}}$ coincides with the decay of $SP(t)$ at short times, as we can 
see in Figs.~\ref{fig:fig03}(b-d). Note, however, that $t^{-\widetilde{D}_{2}}$ coincides better with the decay of $SP(t)$ 
when $\langle k\rangle=4.27$, 5.44 and 6.92, see Figs.~\ref{fig:fig03}(d-f). 
Then, we conclude that $t^{-D_{2}}$ predicts better the decay of $SP(t)$ for small values of $\langle k\rangle$, 
see Figs.~\ref{fig:fig03}(b,c), while $t^{-\widetilde{D}_{2}}$ works better for large values of $\langle k\rangle$,
see Figs.~\ref{fig:fig03}(d-f).
For $\langle k\rangle> 6.92$, see Figs.~\ref{fig:fig03}(g,h), neither $t^{-D_{2}}$ nor $t^{-\widetilde{D}_{2}}$ match the 
decay of $SP(t)$.
Therefore, we can also conclude that the interval of  $\langle k\rangle$ where the decay of $SP(t)$ can be predicted 
by $t^{-D_{2}}$ or $t^{-\widetilde{D}_{2}}$ is quite narrow.

On the other hand, the time-averaged survival probability, also known as the temporal autocorrelation 
function~\cite{KPG92}, has extensively been used to analyze the dynamics of disordered systems at the 
mobility edge~\cite{KPG92,HS94,SBK06}. More recently, it has been applied to the study of the 
many-body localization transition in the 1D Heisenberg model~\cite{TS15}. In fact, it has been observed 
that the dimension $\widetilde{D_{2}}$  directly governs the decay of the time-averaged survival probability, 
which follows a power-law proportional to $t^{-\widetilde{D_{2}}}$. Due to its relevance, the analysis of this 
quantity seems pertinent for our study. 

The time-averaged survival probability is defined as:
\begin{equation}
C(t) = \frac{1}{t}\int_{0}^{t}SP(\tau)d\tau .
\end{equation} 
Thus, in Figs.~\ref{fig:fig04}(a-h) we present $C(t)$ for the same values of the average degree $\langle k\rangle$ 
and network sizes $n$ reported in Fig.~\ref{fig:fig03}. In every panel, we also plot $t^{-\widetilde{D_{2}}}$ as red 
dashed lines, illustrating that the decay of $C(t)$ is well described by this power-law. 
Notably, the best agreement between $C(t)$ and $t^{-\widetilde{D_{2}}}$ is observed when 
$\langle k\rangle\geq 3.35$. Additionally, for comparison purposes, in Figs.~\ref{fig:fig04}(a-b), we also include 
$t^{-D_{2}}$ as blue dashed lines. As already observed in Figs.~\ref{fig:fig03}(a-b), $t^{-D_{2}}$ is a good indicator 
of the $SP(t)$ decay; interestingly, it also proves to be a good descriptor for $C(t)$ in the same parameter regime.

Finally, for completessnes, in Fig.~\ref{fig:fig04}(g) we present the generalized dimension $\widetilde{D}_{2}$ 
vs.~$\langle k \rangle$. This plot illustrates the asymptotic transition to $\widetilde{D}_{2}=1$ (horizontal blue 
dashed line) as $\langle k\rangle$ increases. We observe that the metallic regime ($\widetilde{D}_{2}\rightarrow 1$) 
is reached already when $\langle k\rangle\approx 10$, in agreement with the analysis of the 
relative depth of the correlation hole; see the next Section.

\subsection{Correlation hole}
\label{subsec:Correlation hole}

\begin{figure}[ht]
\centering
\includegraphics[width=0.3\textwidth]{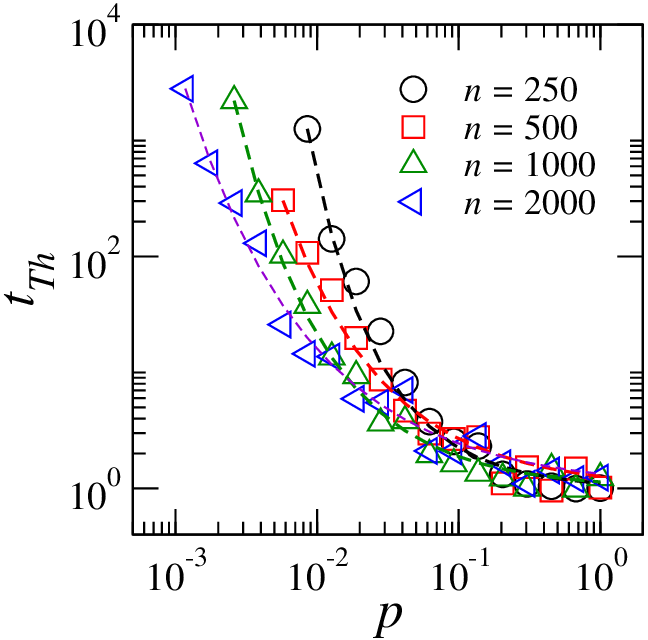}
\caption{
Thouless time $t_{Th}$ of ER random networks of size $n$ as a function of the connection probability 
$p$. Dashed lines correspond to fittings of 
Eq.~(\ref{eq:ThoulessDecay}) to the data with fitting parameters reported in Table~\ref{tab:Thouless}.
}
\label{fig:fig05}
\end{figure}

\begin{figure*}[ht]
\centering
\includegraphics[width=0.75\textwidth]{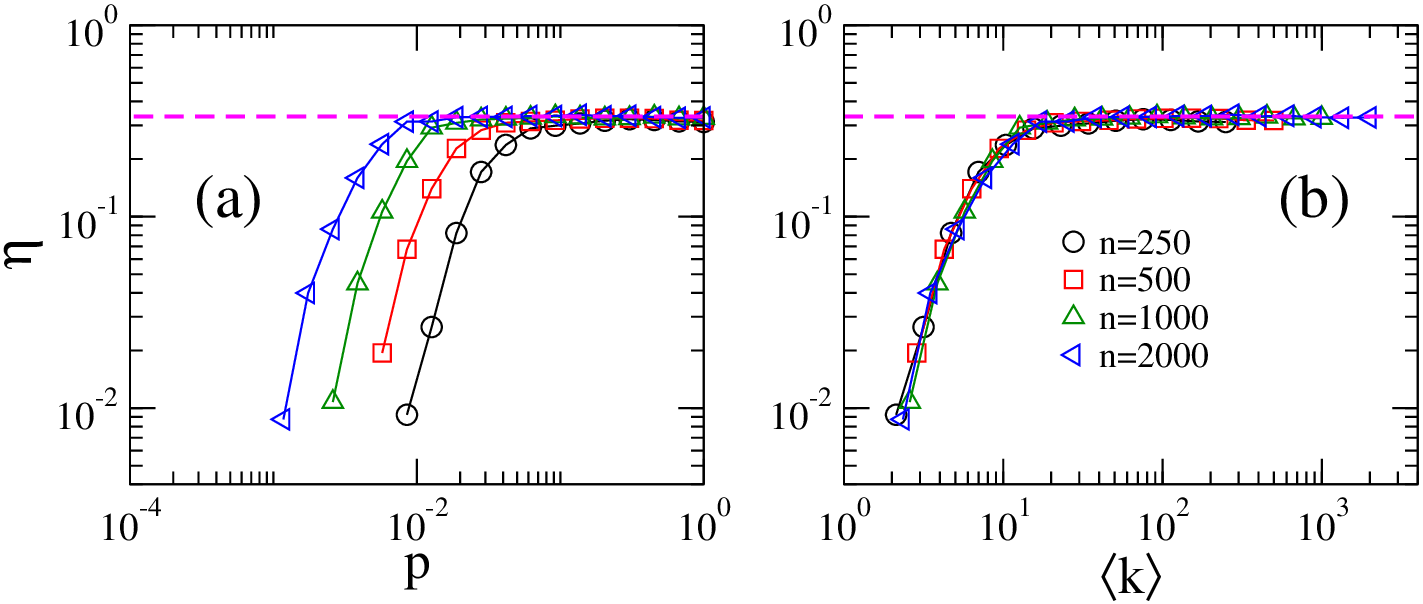}
\caption{
Relative depth of the correlation hole $\eta$ of the survival probability $SP(t)$ for ER random networks 
as a function of (a) the connection probability $p$ and (b) the average degree $\langle k\rangle$. 
Horizontal magenta dashed lines correspond to $\eta=1/3$, the GOE limit.
}
\label{fig:fig06}
\end{figure*}

\begin{table}[t]
\setlength{\tabcolsep}{10pt}
    \renewcommand{\arraystretch}{1.5}
    \centering
    \begin{tabular}{|c |c |c |c |c |} \hline
       \multicolumn{3}{|c|}{$\bf{t_{Th}}$} \\ \hline  \hline
    $n$ & $A$ & $B$ \\ \hline
    250 & 0.105 & 0.886\\ \hline
    500 & 0.239 & 0.615 \\ \hline
    1000 & 0.131 & 0.685 \\ \hline 
    2000 & 0.287 & 0.492 \\ \hline
    \end{tabular}
    \caption{
    Values of the parameters $A$ and $B$ in Eq.~(\ref{eq:ThoulessDecay}) as extracted from fittings to the curves $t_{Th}$ vs.~$p$ of Fig.~\ref{fig:fig05}.
    }
    \label{tab:Thouless}
\end{table}

From Fig.~\ref{fig:fig02}, note that the decay of the $SP(t)$ persists until it reaches its minimum value at a time which 
is known as the Thouless time, $t_{Th}$; see e.g.~\cite{STS19}. Indeed, the minimum value of the $SP(t)$ is also the
bottom of the correlation hole.

Then, in Fig.~\ref{fig:fig05} we report $t_{Th}$ of ER random networks of size $n$ as a function of $p$.
We observe that for a given value of $p$, the larger the network size the smaller the Thouless time.
Also, for a given $n$, we observe the exponential decay of $t_{Th}$ with $p$ which is well described by
    \begin{equation}
    t_{Th} \approx e^{Ap^{-B}},
    \label{eq:ThoulessDecay}
    \end{equation}
    where $A$ and $B$ are fitting constants; see the dashed lines in Fig.~\ref{fig:fig05}
    
%    Additionally, as we dilute the network ($p\rightarrow 0$,$\langle k\rangle\rightarrow 0$),
% the correlation hole vanishes according to the delocalization transition $\langle k\rangle$ value, reported around $\langle k\rangle\approx 1.4$ and we report here $\langle k\rangle\geq 4$~\cite{MAMRP15,EE92,E92,FM91,PV06},  and after its disappearance the minimum of the SP is recovered around the $t_{Th}$ of the GOE~\cite{STS19} (not shown here),
%  $t_{Th}^{GOE}=(3/\pi)^{1/4}\approx 1$, indicating the end of the quasi-diffusive behavior of the system and the onset to localization.  

Additionally, the relative depth of the correlation hole has been used to detect the integrability-to-chaos 
transition in disordered systems and more recently to detect the many-body localization phase in 
disordered interacting systems~\cite{TS17}. Therefore, the study of this quantity is relevant to our study. 

The relative depth of the correlation holes is defined as
\begin{equation}
\eta = \frac{\overline{SP}- SP_{min}}{\overline{SP}}.
\label{eq:RelativeDepth}
\end{equation}
Here, $\overline{SP}$ is the saturation value of the $SP(t)$, and $SP_{min}$ is the minimum value of the
$SP(t)$. In the case of the GOE, $\overline{SP}\approx 3/n$ and  $SP_{min}\approx 2/n$. Then, we 
expect to observe a transtion from $\eta\rightarrow 0$ when $p\rightarrow  0$ to $\eta\rightarrow 1/3$ when 
$p\rightarrow 1$.

Thus, in Fig.~\ref{fig:fig06}(a) we present $\eta$ for ER random networks as a function of $p$. From this figure we 
observe, as expected, that $\eta$
approaches 1/3 for increasing $p$. Moreover, the larger the network size $n$, the smaller the value of $p$ needed
for $\eta$ to approach the GOE regime; that is, the curves $\eta$ vs. $p$ are displaced to the left in the $p$-axis 
for increasing $n$.
Now, when we plot $\eta$ as a function of $\langle k\rangle$, see Fig.~\ref{fig:fig04}(b), all the curves fall one on 
top of the other indicating that $\langle k\rangle$ is the scaling parameter of $\eta$. 
This is in agreement with previous studies where it has been shown that $\langle k\rangle$ scales topological~\cite{AHMS20}, 
spectral~\cite{MAMRP15}, and transport~\cite{MAM13} properties of ER random networks.
Moreover, at $\langle k\rangle \approx 10$, the relative depth of the correlation hole reaches the GOE 
regime~\cite{MAMRP15,TFM20}, depicted by the magenta dashed line at $\eta=1/3$. This means that regardless 
of the parameter combination $(n,p)$, the ER random networks are in the metallic regime for $\langle k\rangle\ge 10$. 

%Since, the curves overlap we can find a heuristic expression that describes the $\eta$  behavior. In this case, we found that
%\begin{equation}
%\eta\approx \frac{1}{3}\left[ 1-\exp\left\lbrace-C\langle k\rangle^{D}\right\rbrace\right]
%\label{eq:ScaledDepth}
%\end{equation}
%describes the general behavior of $\eta$ when $\langle k\rangle>3$ as observed in Fig.~\ref{fig:fig04}(b). Here,  $C$ and $D$ are fitting constants; see Fig.\ref{fig:fig04}(b).

\subsection{Saturation value of the SP}
\label{subsec:Saturation value of the SP}

We also note from Fig.~\ref{fig:fig02}(a) that the saturation values of 
the $SP(t)$ are smaller the larger $p$ is, while the $SP(t)$ reaches its saturation value faster for increasing $p$.
Also, as the networks become more 
connected, the saturation values of the $SP(t)$ approach the GOE prediction of $3/n$, see the insets in 
Figs.~\ref{fig:fig02}(b,c).
     
\begin{figure}[ht]
\centering
\includegraphics[width=0.3\textwidth]{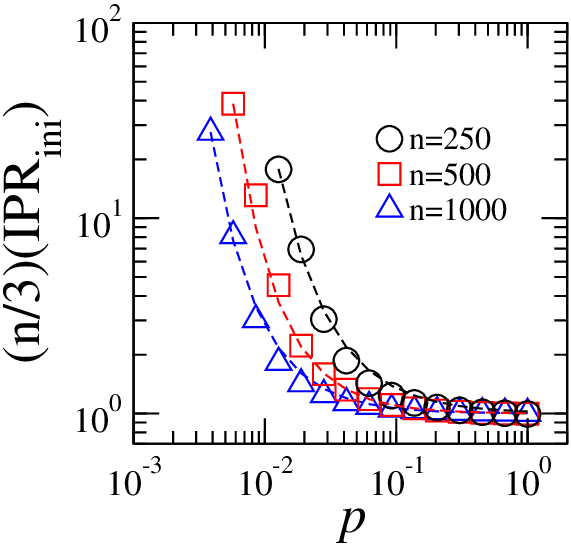}
\caption{
Inverse participation ratio of the initial state $\mathrm{IPR}_{ini}$ (divided by $3/n$) as a 
function of the connection probability $p$. Dashed lines are fittings of 
Eq.~(\ref{eq:IPRiniDecay}) to the data with fitting parameters reported in Table~\ref{tab:IPRini}.
 }
\label{fig:fig07}
\end{figure}

In Fig.~\ref{fig:fig07} we plot the inverse participation ratio of the initial state $\mathrm{IPR}_{ini}$ (divided by 
$3/n$, the GOE saturation value) as a function of $p$ for three different network sizes: $n=250$, 500 and 1000 
(represented by different colors). Since the $\mathrm{IPR}_{ini}$ is the saturation value of the $SP(t)$, the ratio 
$\mathrm{IPR}_{ini}/(3/n)$ approaches 1 as $p\rightarrow 1$, as shown in Fig.~\ref{fig:fig07}. 
Interestingly, we found that the decay of $\mathrm{IPR}_{ini}/(3/n)$ is dictated by the heuristic expression:
\begin{equation}
\mathrm{IPR}_{ini}\approx \frac{3}{n}e^{Ap^{-B}},
\label{eq:IPRiniDecay}
\end{equation} 
where $A$ and $B$ are fitting constants.
Indeed, the dashed lines in Fig.~\ref{fig:fig07} correspond to fittings of Eq. (\ref{eq:IPRiniDecay}) to the data
with the fitting constants reported in Table~\ref{tab:IPRini}.

\begin{table}[h]
\setlength{\tabcolsep}{10pt}
    \renewcommand{\arraystretch}{1.5}
    \centering
    \begin{tabular}{|c |c |c |} \hline
    \multicolumn{3}{|c|}{$\mathrm{\bf{IPR_{ini}}}$} \\ \hline  \hline
    $n$ & $A$ & $B$  \\ \hline
    250 & 0.02408 & 1.096 \\ \hline
    500 & 0.00475 & 1.287  \\ \hline
    1000 & 0.00368 & 1.224  \\ \hline 
    \end{tabular}
    \caption{
    Values of the parameters $A$ and $B$ in Eq.~(\ref{eq:IPRiniDecay}) 
    as extracted from fittings to the curves $\mathrm{IPR_{ini}}$ vs.~$p$ 
    of Fig.~\ref{fig:fig07}.
    }
    \label{tab:IPRini}
\end{table}

\section{Conclusions}\label{Sec:Conclusions}

In this work, within a random matrix theory (RMT) approach, we have analyzed the survival probability 
$SP(t)$ of Erd\"os-Renyi (ER) random networks in the crossover from isolated nodes to fully connected 
networks. Note that the crossover from isolated nodes to fully connected networks corresponds to 
the crossover from localized to extended eigenstates of the corresponding randomly weighted adjacency 
matrices. We recall that the ER model is defined as $n$ nodes connected randomly with probability $p$.

We have observed that the time evolution of the SP of a delta-like excitation in ER random networks 
shows the standard panorama reported for RMT models:
The SP displays an initial fast decay followed by a power-law decay, then it reaches a minimum value
(at a time which is known as the Thouless time $t_{Th}$) before saturation; see Fig.~\ref{fig:fig02}.

For the fast decay of the SP we were able to write down an expression that depends on the average degree 
$\langle k\rangle\approx np$ only, see Eq.~(\ref{SPfast}). Also, we have observed that once 
$\langle k\rangle$ is fixed, the curves of $SP(t)$, as well as its time-average $C(t)$, coincide for different
network sizes in the power-law decay regime; see Figs.~\ref{fig:fig03} and~\ref{fig:fig04}.
Moreover, we showed that the power-law decay of $SP(t)$ and $C(t)$ is well approximated by $t^{-D_{2}}$ 
and $t^{-\widetilde{D}_{2}}$, respectively. Here, $D_2$ and $\widetilde{D}_2$ are the correlation dimension 
of the eigenstates of the ER random networks and the correlation dimension associated with the initial state,
respectively.

In addition, we showed that the relative depth of the correlation hole of the SP scales with $\langle k\rangle$ 
(see Fig.~\ref{fig:fig06}), while both, $t_{Th}$ and the IPR of the initial state, decay exponentially with $p$;
see Figs.~\ref{fig:fig05} and~\ref{fig:fig07}.
Finally, we provided strong evidence of multifractality of the eigenstates of the randomly weighted adjacency 
matrices of ER networks; see Figs.~\ref{fig:figA1} and~\ref{fig:figA2}. 

We hope that our study may motivate further research on the application of RMT techniques to the study of 
dynamical properties of random network models.

\appendix
\section{Multifractal dimensions of eigenstates}
\label{append:Multifractal dimensions of eigenstates}
\renewcommand\thefigure{\thesection.\arabic{figure}}
\setcounter{figure}{0}

In Sec.~\ref{subsec:Decay of the SP} we have shown that the decay of the survival probability $SP(t)$ 
and that of the time-averaged survival probability $C(t)$ of ER random networks are governed by the
power-laws $t^{-D_2}$ and $t^{-\widetilde{D}_2}$, respectively. Here, $D_2$ is the correlation dimension
of the eigenstates of the adjacency matrix of the ER random networks while $\widetilde{D}_2$ is the
correlation dimension of the initial state. 
Thus, in this Appendix we give details of the calculation of $D_2$ and $\widetilde{D}_2$ but we also 
report the generalized dimensions $D_q$.

Disordered systems that manifest the Anderson transition, a metal-to-insulator transition
driven by the disorder amplitude, exhibit a number of critical properties. 
One well established signature of the Anderson transition is that the eigenstates
of the corresponding disordered systems are multifractal objects at and in a small neighborhood of 
the critical transition point~\cite{EM08}. Lattice models, such as the 3D Anderson 
tight-binding model, the power-law random banded matrix (PRBM) 
model, and critical random matrix ensembles, have been extensively used to study the Anderson 
transition~\cite{MAV12,EM08,MFDQS96}.
In addition, multifractal eigenstates at the critical point have been reported for random graph 
models~\cite{GGGMDL17} as well as for a diluted version of the PRBM~\cite{VMR19} which represents
ER-like random networks with long-range interactions.

The multifractality of the eigenstates $\Psi^\alpha$ is characterized by the generalized dimension $D_q$,
which is computed from the size scaling of the average inverse participation ratio as~\cite{MAV12,BG12}
\begin{equation}
\left\langle \mathrm{IPR}(q)\right\rangle \equiv
\left\langle {\sum_{i=1}^{n}\left|\Psi_i^\alpha\right|^{2q}} \right\rangle_\alpha \propto n^{-(q-1)D_q}, \quad q\neq 1.
\label{eq:Dq}
\end{equation} 
While the information dimension $D_1$ is computed from the size scaling of the average Shanon entropy
\begin{equation}
\left\langle S \right\rangle \equiv
\left\langle -\sum_{i=1}^{n}\left|\Psi_i^\alpha\right|^{2}\ln{\left|\Psi_i^\alpha\right|^2}\right\rangle_\alpha \propto D_1 \ln n.
\label{eq:D1}
\end{equation} 
When computing $D_q$ for the eigenstates of a disordered system, $D_q=d$ and $D_q=0$ correspond to 
fully delocalized states and localized states, respectively. Here, $d$ is the embedding dimension of the 
system. While $0<D_{q}<d$ indicates multifractal states, 
meaning that such states do not extend over the entire available configuration space.

\begin{figure*}[ht]
\centering
\includegraphics[width=0.65\textwidth]{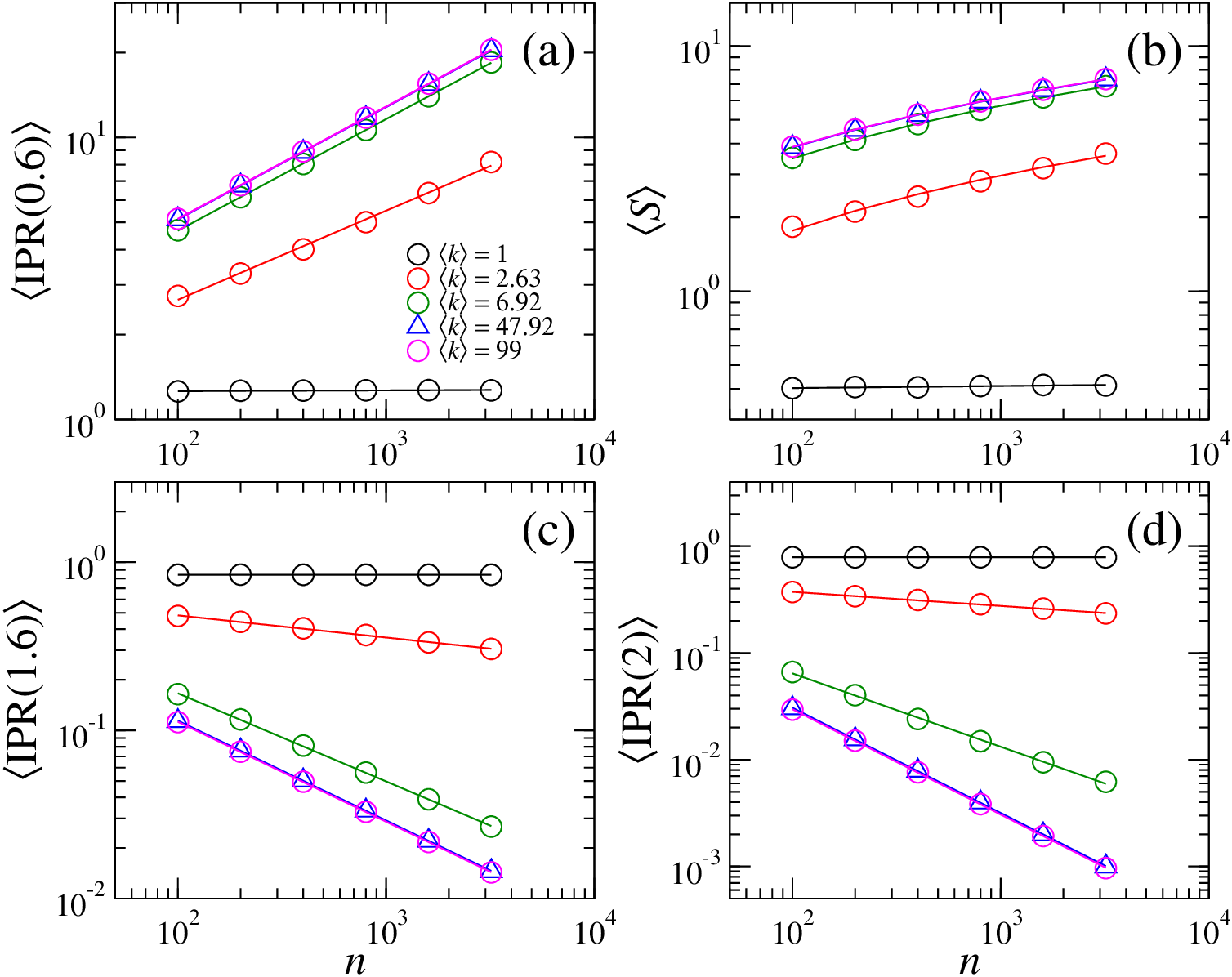}
\caption{
Average inverse participation ratio $\left\langle \mathrm{IPR}(q)\right\rangle$ of ER random networks as a 
function of the size $n$ for (a) $q=0.6$, (c) $q=1.6$, and (d) $q=2$. In panel (d), the average Shannon entropy 
$\left\langle S \right\rangle$ of ER random networks as a function of $n$ is shown.
Several values of the average degree $\langle k\rangle$ are reported in each panel, as indicated in panel (a).
The averages are taken over $20\%$ of the eigenstates at the center of the band and over $10^{6}/(0.2 n)$ 
random network realizations.}
\label{fig:figA1}
\end{figure*}

\begin{figure*}[ht]
\centering
\includegraphics[width=0.65\textwidth]{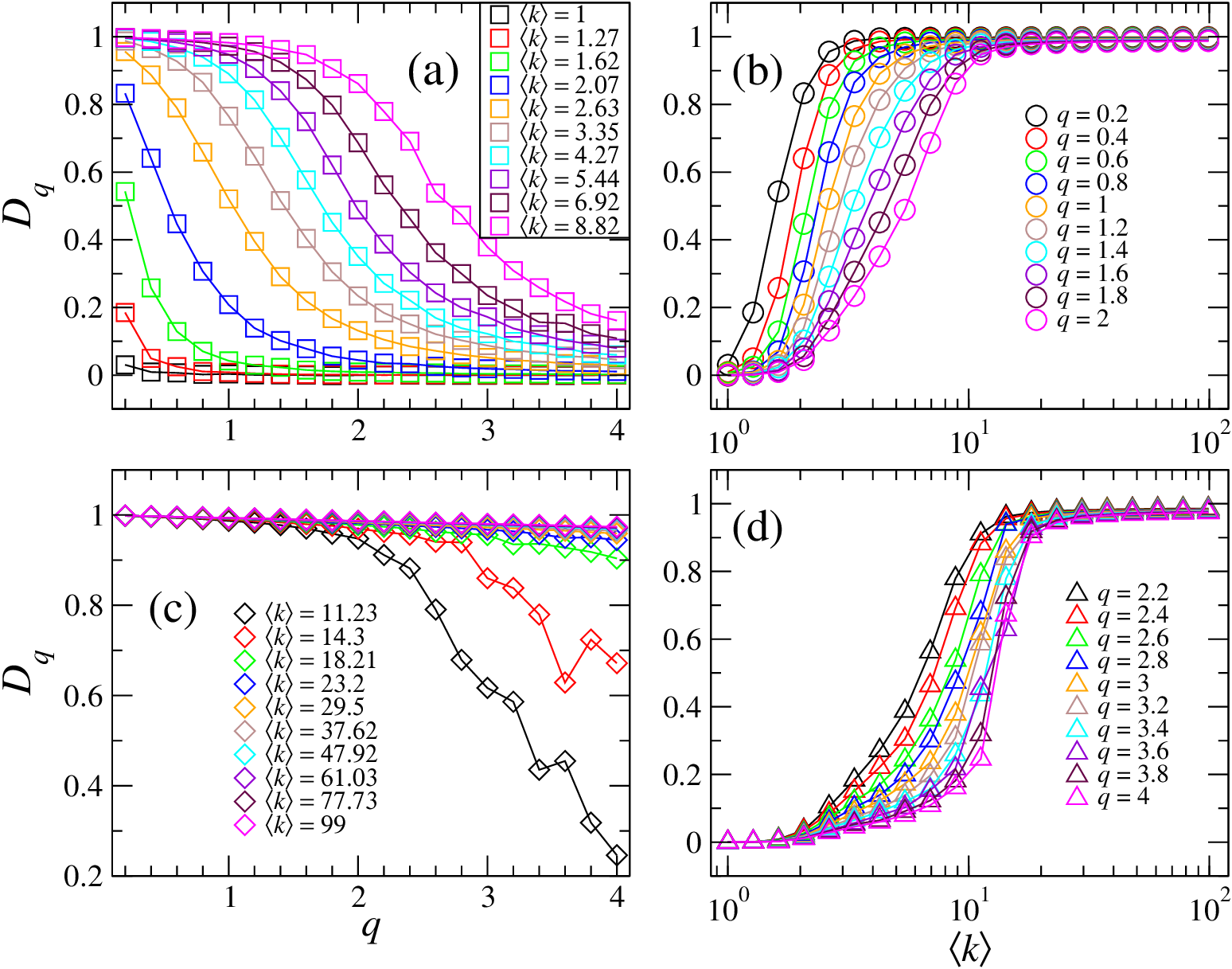}
\caption{
Multifractal dimensions $D_{q}$ of ER random networks as a function of (a,c) $q$ and (b,d)  
$\langle k\rangle$ for several values of $\langle k\rangle$ and $q$, respectively.
Error bars are not shown since they are smaller than symbol size.
}
\label{fig:figA2}
\end{figure*}

Here, since we deal with ER random networks represented by randomly weighted adjacency matrices
which are members of a diluted GOE that interpolates between the Poisson ensemble when $p\to 0$ 
and the GOE when $p\to 1$, we expect to observe a transition from $D_{q}=0$ to $D_{q}=1$ when
increasing $\langle k\rangle$ from $\langle k\rangle\approx 0$ to $\langle k\rangle\approx n-1$.

In Fig.~\ref{fig:figA1} we present the $\left\langle \mathrm{IPR}(q)\right\rangle$ of ER random networks 
as a function of $n$, in log-log scale, for $q=0.6$, 1, and 2; as well as $\left\langle S \right\rangle$ 
vs.~$n$. In each panel we show curves for different values of $\langle k\rangle$. From Fig.~\ref{fig:figA1} 
we observe that the slopes (in log-log scale) of the curves of $\left\langle \mathrm{IPR}(q)\right\rangle$
vs.~$n$ and of $\left\langle S \right\rangle$ vs.~$n$, which coincide with $D_q$ and $D_1$, respectively, 
are approximately zero for small $\langle k\rangle$ while they approach one for large $\langle k\rangle$.
This indicates a transition from an insulating regime (localized eigenstates) to a metallic regime (extended
eigenstates) by increasing $\langle k\rangle$.
Remarkably, the slopes between zero and one of the curves $\left\langle \mathrm{IPR}(q)\right\rangle$
vs.~$n$ and $\left\langle S \right\rangle$ vs.~$n$, observed in Fig.~\ref{fig:figA1} for intermediate values 
of $\langle k\rangle$, indicate the existence of multifractal eigenstates.

Then, in Fig.~\ref{fig:figA2} we present $D_{q}$ of ER random networks as a function of $q$ (see left
panels) and also as a function of $\langle k\rangle$ (see right panels).
The values of $D_q$ reported in Fig.~\ref{fig:figA2} are obtained from fittings of the data 
$\left\langle \mathrm{IPR}(q)\right\rangle$ vs.~$n$ and $\left\langle S \right\rangle$ vs.~$n$ with
Eqs.~(\ref{eq:Dq}) and~(\ref{eq:D1}), respectively. In Fig.~\ref{fig:figA2} error bars are not shown since 
they are smaller than symbol size.

From Fig.~\ref{fig:figA2} we note a clear multifractal behavior for the eigenstates of the weighted
adjacency matrices of ER random networks: $D_q\ne D_{q'}$ for $q\ne q'$. 
Also, we clearly observe the metal-insulator transition as a function of $\langle k\rangle$:
$D_q\approx 1$ for large $\langle k\rangle$ and $D_q\approx 0$ for small $\langle k\rangle$.

Finally, we mention that we computed $\widetilde{D}_2$ from the 
scaling of the inverse participation ratio of the initial state, that is~\cite{TS15}
\begin{equation}
\left\langle \mathrm{IPR(2)}\right\rangle_{ini} = \left\langle \sum_{\beta}\left|C_{ini}^{\beta}\right|^{4}\right\rangle \sim n^{-\widetilde{D}_{2}}.
\label{eq:IPR_D2tilde}
\end{equation}
In our analysis we computed $\left\langle \mathrm{IPR(2)}\right\rangle_{ini}$ for networks of sizes $n = 100$, 
200, 400, 800, 1600, and 3200 over an eigenstate window of $20\%$ of the matrix size around $\ket{\phi_{ini}}$.

\section*{Acknowledgements}
The authors thank Lea F. Santos for valuable discussions and feedback on an early draft of this manuscript.\\
K.P.-M. thanks support from SECIHTI (Postdoctoral Fellowship CVU 1008060), Mexico.\\
J.A.M.-B. thanks support from SECIHTI (Grant No.~CBF-2025-I-2236), Mexico.

\bibliographystyle{sorsamp.bib}

\end{document}